# Integration I(d) of Nonstationary Time Series
*Stationary and nonstationary increments*

Joseph L. McCauley, Kevin E. Bassler[+], and Gemunu H. Gunaratne[++]

Physics Department
University of Houston
Houston, Tx. 77204-5005
jmccauley@uh.edu

[+]Texas Center for Superconductivity
University of Houston
Houston, Texas

[++]Institute of Fundamental Studies
Kandy, Sri Lanka

**Abstract**
The method of cointegration in regression analysis is based on an assumption of stationary increments. Stationary increments with fixed time lag are called 'integration I(d)'. A class of regression models where cointegration works was identified by Granger and yields the ergodic behavior required for equilibrium expectations in standard economics. Detrended finance market returns are martingales, and martingales do not satisfy regression equations. We extend the standard discussion to discover the class of detrended processes beyond standard regression models that satisfy integration I(d). In the language of econometrics, the models of interest are 'unit root' models, meaning martingales. Typical martingales do not have stationary increments, and those that do generally do not admit ergodicity. Our analysis leads us to comment on the lack of equilibrium observed earlier between FX rates and relative price levels.

## 1. Definition of Integration I(d)

First, we establish our terminology and notation. Given a stochastic process x(t) or a time series realization of a process, economists call a point x(t) a level, and the increment/displacement x(t,-T)=x(t)-x(t-T) is called a level difference, or just difference. With logarithmic variables as in finance or discussions of the money supply, x(t,T)=ln(p(t+T)/p(t)) is the log increment of a price, whereas the underlying stochastic process is defined by the random variable x(t)=ln(p(t)/$p_c$) where $p_c$ is a reference price that can be identified as 'value' [1].

There is exactly one, single time scale t in a random variable x(t). There are *two* time scales in an *increment*, the time t and a time lag T. In economics the time lag is typically fixed at T=1 period [2], e.g. 1 period can mean 1 quarter, and the idea that T is a time variable is completely ignored. This may lead to confusion, because the time lag T may increase without bound and increments x(t,T) of a nonstationary process x(t) *cannot* be understood as 'stationary processes' as T is increased even if the increments are stationary. That is, stationarity of increments is very different from stationarity of a process. In our recent FX data analysis [3] we studied increments of detrended returns with T=10 min. and in the accompanying theory it was made clear that T is a variable. Summarizing, in agreement with mathematics terminology and our recent papers, we will denote as x(t) the (stochastic) process or point in a time series, and x(t,T) is an increment of the process. *The central question for us is: what is the class of a stochastic processes {x(t)} admitting stationary increments.* By a stationary increment is meant that x(t,T)=x(0,T) 'in distribution'. When T is held fixed, e.g., at T=1, then we will make contact with economists' language and expectations. **Note that stationarity of an** *increment* **means only that the 1-point increment distribution is independent of t, it does not mean that 2-point and higher order distributions satisfy any stationarity condition at all, and generally they do not. In contrast, even weakly stationary** *processes*[1] **x(t) demand that both the 1-point and 2-point distributions are time translationally invariant. The latter is a necessary but not sufficient condition that ergodicity may occur.**

The belief in econometrics and macroeconomic data analysis is that increments may be stationary, with fixed T, even the process is not [2,4]. Stated explicitly, if the process x is nonstationary then form the first difference x(t,-T)=x(t)-x(t-T). If the first difference x(t,-T) is nonstationary with T fixed, then one may study the second difference $x(t-T,-T) = x(t) - 2x(t-T) + x(t-2T)$, and so on until either stationarity is found or the chase is abandoned. If the process x(t) is already stationary then it's called I(0). If the process x(t) is nonstationary but the first difference x(t,-T) is stationary for fixed T then the process is called integrated of order 1, or simply I(1). If neither x nor the first difference is stationary but the second difference is stationary for fixed T, then the process is called I(2). According to Johansen [5,6] the above prescription is not entirely correct: it's possible to construct a two variable regression model based on special assumptions about the noise where both the processes x,y and differences x(t,-1), y(t,-1) are nonstationary but there is still a stationary linear combination αx(t)+βy(t). This is an aside. We'll explain why the models of concern to us, including finance market returns, are not amenable to regression analysis.

---

[1] Stationary processes are defined in part 4.

***By 'noise' we mean any drift free stochastic process.*** Typical examples of noise are the Wiener process, white noise, statistically independent nonstationary noise, i.i.d. noise, drift free stationary processes, martingales, fractional Brownian motion, and the correlated noise of near-equilibrium statistical physics. "Noise" therefore implies nothing whatsoever about correlations of any sort, only that the process is drift free and stochastic. By drift, we mean an additive term A such that the process has the form x(t)=A+noise. The drift may be a functional of x, and the noise may be represented by a stochastic integral with a function of x in the integrand, as in a general Ito process.

*We want to identify the class of drift free processes x(t) for which integration I(1) is possible when T is held fixed.* In theoretical discussions of integration and cointegration either 'white noise' or i.i.d. noise is assumed ad hoc (without empirical basis). In the context of the Granger Representation Theorem, it's been stated that the practitioners of cointegration generally do not worry much about the noise distribution because the cointegration technique is presented primarily as matrix algebra [7]. In cointegration studies the test of the noise distribution generally does not go beyond checking for a (presumably 1-point) Gaussian distribution [8]. It's also known that a change of time variable is sometimes adequate to transform nonstationary differences to stationary ones but is generally inadequate [8], and we will explain in part 5 below why such a time transformation generally cannot be found.

The tradition in macroeconomics is to postulate the noise properties in as simple a way as possible ("i.i.d." or "white") instead of discovering the noise distribution from time series analysis, as we did in our finance data analysis [3,9]. We will carefully analyze the distributions of 'white' and i.i.d. noise processes below, and will show that stationary increment martingales include 'white noise' and provide a more general and empirically realistic class of noise sources for discussions of I(1) than do i.i.d. sources. First, we present two brief sections summarizing the economists' viewpoint. A vast gap exists between our approach and theirs. Our approach [3,9,10] is based on the deduction of *falsifiable* dynamical models from empirical time series.

## 2. Regression Analysis as a Search for Stationarity

To illustrate regression models [11], consider any macro economic variable x(t) like unemployment, the price level, or the money supply, or an exchange rate. Suppose that by ignoring uncertainty (as in neo-classical economics, e.g.) macroeconomic theory predicts or speculates that $x(t)=\lambda x(t-1)$ should hold, where t-1 is the present time and t is 1 period later. Then the hope is to find that

$$x(t) = \lambda x(t-1) + \varepsilon(t). \qquad (1)$$

where ε(t) is drift free uncorrelated noise with zero mean and finite variance (fat tails in the distribution of ε(t) would kill the last assumption), and λ is a free parameter. This defines what is meant in econometrics by 'white noise' if the variance is taken to be constant. *If* the noise is stationary *then* so is x(t) since there is no drift in (1). Assuming 'white noise' ε(t), we obtain

$$\langle x(t) \rangle = 0$$
$$\langle x^2(t) \rangle = \frac{\sigma_1^2}{1-\lambda^2} \qquad (2)$$

where $\langle \varepsilon^2(t) \rangle = \sigma_1^2 = \text{constant}$ if the drift $\langle x(t) \rangle = \lambda^t x(0)$ has been subtracted from x(t). Stationarity is therefore possible iff. 0<λ<1 (ignoring fat tails), where λ=1 is called a 'unit root' in econometrics. Another macroeconomic way to arrive at (1) is simply to regard it as a regression eqn., and to use standard econometric assumptions to try to test data for its validity. Still a third interpretation is to assert that the monetary authority may try to enforce a rule x(t)=λx(t-1) 'to within error' for the next period t, at present time t-1 based on the presently observed value x(t-1). This interpretation led historically to the interesting question whether governmental policy rules are optimal or even effective [11}.

Econometrics and regression analysis aside, from the standpoint of the theory of stochastic processes the model (1) with λ=1 defines a martingale process *if the noise ε(t) is uncorrelated* [9,10], and martingales are inherently nonstationary [1]. That is, stationary noise in (1) is impossible if λ=1. Indeed, macroeconomists interpret a unit root as necessity to worry about nonstationarity. The simplest martingale is provided by the Wiener process B(t), $\langle B(t) \rangle = B(0) = 0$, $\langle B(t+T)B(t) \rangle = \langle B^2(t) \rangle = t$, so that the process variance $\langle B^2(t) \rangle = t$ is not constant, the process is strongly nonstationary with stationary increments, e.g., $\langle B^2(t,T) \rangle = \langle B^2(0,T) \rangle = T \langle B^2(1) \rangle$.

The process (1) with λ=1 and (x,t)-dependent noise with uncorrelated but nonstationary increments is exactly the model indicated by finance data [3,9]. This is the reason for our focus on martingales. Our point is that the noise "ε(t)" in that case is an increment, albeit not a Wiener process increment.

Continuing with regression analysis, suppose instead of (1) that we consider a twice time-lagged regression eqn.

$$x(t) = \alpha x(t-1) + \beta x(t-2) + \varepsilon(t). \quad (3)$$

We introduce the time shift operator Ly(t)=y(t-1). The noise term in eqn. (3) can then be rewritten as equal to

$$x(t) - \alpha x(t-1) - \beta x(t-2) = (1 - \lambda_1 L)(1 - \lambda_2 L) x(t). \quad (4)$$

Here's the central question in regression analysis: *when is stationary noise possible?* If $0<\lambda_1<1$ then we can set $\lambda_2=0$, but if $\lambda_1=1$ then we have

$$(1-L)(1-\lambda_2 L)x(t) = \varepsilon(t), \quad (5)$$

or with x(t,-T)=x(t)-x(t-T),

$$x(t,-1) = \lambda_2 x(t,-2) + \varepsilon(t), \quad (6)$$

so we see that stationary noise is possible in (3) for $0<\lambda_2<1$ even if eqn. (1) has a unit root. 'Unit root processes' are central: martingales describe detrended financial variables but we advise a direct test for a martingale [3,9] rather than a unit root test.

Ergodic processes are a subclass of stationary processes. Ergodicity here means that time averages converge in probability to ensemble averages. Forming ensemble averages from a single historic time series is illustrated in [3,9]. By i.i.d. is meant statistical independence with stationarity. I.i.d. noise is trivially ergodic, the convergence of time to ensemble averages is provided by the law of large numbers [12]. The stationary process y(t) defined by (1) with $|\lambda|<1$ is ergodic in discrete time [12,13,14]: the pair correlations for a time lag nT, $\langle y(t)y(t+nT)\rangle = R(nT)$, vanish as n goes to infinity. This is the sort of ergodicity assumed in regression analysis models[2]. With a discrete time stationary process the time average always converges, but if the system is not ergodic then the limit is not necessarily the ensemble average [12]. With a nonstationary process there is no possible appeal to time averages, we must construct ensemble averages in order to perform any data analysis at all.

Inadequate distinction is made in regression analysis between noise levels and noise increments (see Kuersteiner [15] for an exception). We will clarify this in part 4 and will point out that the noise "ε(t)" in the regression eqns. like (1,3-5) Is always, by necessity, a *noise increment* ε(t,-T). E.g., in eqn. (1) with λ=1 the noise is exactly a martingale increment x(t,-1)=x(t)-x(t-1). This is made precise in part 4 below.

### 3. The Idea of Cointegration

In macroeconomics, relations between economic variables are expected on the basis of non-empirically based equilibrium argumentation [2,4]. In econometrics, regression analysis is used to try to discover or verify the predicted relationships. Given two time series for two different economic variables x and y, like price levels and the money supply, or FX rates and the relative price levels in two countries, regression analysis assumes a form y=α+βx+ε(t) where the standard assumption in the past was that the noise ε(t) can be treated as a stationary 'error' (nonlinear regression analysis exists in the literature but is irrelevant here). It was realized early on that typical macroeconomic variables x and y (price levels and FX rates, e.g.) are nonstationary [2], and it has been known even longer that regression analysis based on the assumption of stationary noise can easily 'predict' spurious relations between completely unrelated nonstationary variables [2,4]. The assumption of integration I(d) is that with nonstationary random variables x(t), increments x(t,T)=x(t+T)-x(t) are stationary with T fixed, to within a removable drift [7,16]. Cointegration was invented as a generalization of the idea of 'integration I(d)' as a technique for trying to infer both short time (T=1 period) and long time equilibrium (based on ergodicity) relations between nonstationary economic variables via regression analysis [2,4].

---
[2] And is generally absent in the case of stationary increment martingales.

Here's a definition of cointegration quoted literally from the source [4]. Think of macroeconomic variables as the components of a column vector **x**(t). "The components of **x**(t) as said to be co-integrated of order d,b, denoted CI(d,b), if (i) all components of **x** are I(d); (ii) there exists a vector $\alpha \ne 0$ so that $z(t) = \hat{a}x$ is I(d-b), b>0. The vector $\alpha$ is called the co-integrating vector." The 'hat' denotes the transpose, a row vector so that $\hat{a}x$ denotes the scalar product of two vectors. The authors [4] then state that for the case where d=b=1, cointegration would mean that if the components of **x**(t) were all I(1), then the equilibrium error would be I(0), and z(t) will rarely drift far from zero if it has zero mean, and z(t) will often cross the zero line." That is, $\hat{a}x = 0$ is interpreted as an *equilibrium* relationship, and the last sentence above expresses the unproven hope that the stationarity of integration I(d) is the strong stationarity that brings with it the ergodicity of statistical equilibrium ('…will rarely drift far from zero….will often cross the zero line' [4]).

In the Nobel Committee's description of Granger's work [2] it was noted that cointegration failed to exhibit the expected long time equilibrium relationship expected between FX rates and relative price levels in two countries. It was argued therein that cointegration deals with short times, T=1 period, and that short time lags are inadequate to expose the long time equilibrium relations that would follow from ergodicity. We're going to suggest that the real reason for the failure of an equilibrium relation between two financial variables is entirely different, and is not at all due to the restriction to a short time lag T.

In the next two sections we analyze the statistical properties of noise levels and increments. In part 5 we will show that arbitrary stationary increment martingales are the right generalization of 'white noise', and that the assumption of i.i.d. is both unnecessary and too limiting, lack of increments correlations rather than full statistical independence of increments is adequate. This is also practical, because in empirical analysis we generally cannot discover even a 1-point distribution, much less prove that empirical data are i.i.d. even if they would be!

## 4. The Distributions of I.I.D. and 'White Noise' Processes and Increments

### 4.1 Stationary Processes Defined

It's necessary to avoid confusing stationary increments with a stationary process. Stationary increments are defined in [12,13,17,18] although that term is not used in the translation of Gnedenko [12]. A stationary process is one where a time independent 1-point distribution exists and is normalizable (guaranteeing time independence of the mean and variance of the process), and all higher order densities $f_n$, $n \geq 2$, are time translationally invariant as well, $f_n(x_n,t_n+T;...;x_1,t_1+T) = f_n(x_n,t_n;...;x_1,t_1)$. The pair correlations following from $f_2$ are therefore time translationally invariant, $\langle x(t+T)x(t)\rangle = \langle x(T)x(0)\rangle$, violating the martingale condition $\langle x(t+T)x(t)\rangle = \langle x^2(t)\rangle$. Ergodic processes[3] are a subset of stationary processes [12,13,14].

In 'wider form' or weak stationarity, time translational invariance of only two densities $f_n$ for n=1,2 is sufficient: the mean and variance are constants, and $\langle x(t+T)x(t)\rangle = \langle x(T)x(0)\rangle$. We've pointed out above that stationarity of *increments* is a far weaker condition than either strict or weak stationarity of a process. In the former case, only the increment density need be independent of t, nothing at all is required of the pair correlations (fractional Brownian motion has long time increment correlations and has stationary increments). In particular, increment stationarity requires that the increment density f(z,t,T) depends on t alone, places no restriction whatsoever on pair correlations, and certainly does not imply an i.i.d. distribution.

## 4.2 I.i.d. Noise

Because econometrics and macroeconomics nearly always assumes either 'white' or i.i.d. noise, we now analyze the statistical properties of both. We will show that an i.i.d. (drift free, statistically independent, identically distributed) noise process ε(t) generally cannot generate stationary increments, therefore is not I(d). In contrast with standard econometrics practice we'll distinguish carefully between the distributions of noise levels and noise level differences. We want to show that the noise "ε(t)" in regression analysis is always a noise increment, and that white noise, not i.i.d. noise, is the correct basis for relaxing the restrictions imposed in regression analysis to include the empirically observed noise of finance markets.

---

[3] With an ergodic process, long time averages converge in probability, or in mean square, to ensemble averages.

For the Wiener process B(t), the simplest martingale with stationary increments, one obtains from $f(x,t,t+T) = \int dxdy f_2(y,t+T;x,t)\delta(z-y+x)$ that $f(z,t,t+T) = p_2(z,T|0,0)$ where $p_2$ is the transition density of the Wiener process, $f_2(y,t+T;x,t) = p_2(y,t+T|x,t)f_1(x,t)$. Although the Wiener process B(t) is not i.i.d, the Wiener *increments* B(t,T) are i.i.d. in the following precise sense *iff.* we restrict our considerations to T=constant: the Wiener process B(t) is Markovian

$$f_n(x_n,t_n;x_{n-1},t_{n-1};...;x_1,t_1)/f_1(x_1,t_1) = p_2(x_n,t_n|x_{n-1},t_{n-1})p_2(x_{n-1},t_{n-1}|x_{n-2},t_{n-2})...p_2(x_2,t_2|x_1,t_1)$$
. (7)

If we combine the Markov condition with the time and space translational invariance of the Wiener process $p_2(x_n,t_n|x_{n-1},t_{n-1}) = p_2(x_n - x_{n-1}, t_n - t_{n-1}|0,0)$, then this casts (7) into the form of a condition for i.i.d. *increments* where the 1-point density of increments is exactly $f(z,t,t+T) = p_2(z,T|0,0)$[3]. That is, the random walk is not i.i.d., random walk increments are not i.i.d. (with T varying stationarity doesn't hold), but with T=constant we may treat random walk increments as i.i.d. *The basic example of i.i.d. increments is not a stationary process*. Now for the main point.

First, for I(1) noise x(t), all we need is that x(t,1) and x(0,1) have the same 1-point distribution. *Full statistical*

---

[3] Without both time and space translational invariance, one cannot obtain an i.i.d. distribution for increments from a Markov condition.

*independence is unnecessary, all we need is the condition of uncorrelated, stationary increments*. Second, deducing i.i.d. conditions from empirical data would be impossible, the best that can be hoped for empirically is to test for vanishing increment autocorrelations. Therefore, the i.i.d. assumption can and should be replaced by the much more general condition of a martingale with stationary increments (martingales are Markovian iff. there is no finite memory in the transition density [9]). All drift free Markov processes are included in martingales, e.g. This observation opens up macroeconomics to the consideration of nontrivial dynamics not studied in regression analysis. And, we can show how such data can and should be analyzed by forming ensembles from a single historic time series [3].

### 4.3 'White Noise' in Econometrics consists of Wiener Process Increments with fixed time lag T

To back up our claim of the importance of martingales for integration I(d), here's the simplest I(1) case presented in the literature [19]. Consider the random walk on the line, the Wiener process B(t). The Wiener process is a martingale, $\langle B(t)B(t-T)\rangle = \langle B^2(t-T)\rangle$, with stationary increments

*The increment process (21) has been labeled as "white noise" in econometrics [19].* In the economists' 'white noise' the increments are uncorrelated, $\langle B(t,1)B(t,-1)\rangle = 0$ and have constant variance $\langle B^2(1)\rangle = 1$, nothing more. In other words, what has been called 'white noise' in econometrics is actually the simplest martingale with stationary increments and fixed time lag. *We can there forget about both 'white noise' and i.i.d. noise and focus instead on the much more general case of stationary increment martingales in order to define Integration I(d) for drift free stochastic processes.*

With T fixed, any stationary increment martingale is I(1). For martingales I(d) with d≥2, is superfluous. Furthermore, even with stationary increments and T =1, x(t,T)=x(0,1), there is no ergodicity (excepting the Wiener process) because martingale correlations are nonstationary, and without the time and 'space' (x) translational invariance of the Wiener process then even a Markovian martingale does not generate i.i.d. increments. This means that time averages of stationary increments generally do not converge to the ensemble average values $\langle x(t,T) \rangle = \langle x(0,T) \rangle = 0$, $\langle x^2(t,T) \rangle = \langle x^2(0,T) \rangle = \langle x^2(T) \rangle - \langle x^2(0) \rangle$, neither for T=1 nor for long time lags T. Because martingale serial correlations are nonstationary there is no power spectrum for x(t), and there is also no basis whatsoever for assuming a power spectrum for x(t,-1). Martingales are discussed in detail in references [9,10].

## 5. Martingale Noise

An Ito process is generated locally by a drift term plus a martingale, $dx = R(x,t)dt + b(x,t)dB(t)$, where B(t) is the Wiener process and b is a 'nonanticipating function' of the random variable x [20]. Ito processes in general, and local martingales

$$dx = b(x,t)dB(t) \qquad (8)$$

in particular became of central importance in mathematical finance with the formulation of the EMH [21] and the success of the Black-Scholes model [22]. We've deduced above how to generalize the noise assumed in regression analysis to include martingales, providing a bridge between ideas in finance theory and econometrics.

Examples of 1-point densities calculated from specific diffusion coefficients $b^2(x,t) = D(x,t)$ are given in [23,24]. Setting $b^2(x,t)=D(x,t)=D(t)$, or $b(x,t)=x$ (lognormal process), in (8) generates two different martingales, but each is equivalent to the Wiener process by a specific coordinate transformation. Setting $D(x,t) = |t|^{2H-1}(1+|x|/|t|^H)$ with H≈.35 in (8) describes a nontrivial martingale observed during one time interval of intraday FX trading [1,3]. This martingale is topologically inequivalent to the Wiener process [10].

## 5.1 Stationary Increment Martingales

Consider first the class of all drift-free nonstationary processes with uncorrelated stationary increments. For this class the 1-point density of the random variable x is nonstationary, and the increments x(t,T) and x(0,T) have the same nonstationary 1-point distribution as a function of the time lag T as has the 1-point distribution of the process x(t) as a function of the starting time t. This is the meaning of stationary increments. Examples are (i) the Wiener process B(t) (where $\langle B^2(t) \rangle = t$) where the increments are uncorrelated, and (ii) any martingale with variance linear in t, $\langle x^2(t) \rangle = t \langle x^2(1) \rangle$, where x(0)=0 is required, and the increments are always uncorrelated but generally are not i.i.d. due to lack of translational invariance even if the process is Markovian. An example where the increments are stationary but strongly correlated is provided by fractional Brownian motion, where $\langle x^2(t) \rangle = t^H \langle x^2(1) \rangle$, 0<H<1, and in addition the increments are strongly correlated if H≠1/2. For class (ii), which includes class (i), if we study increments with fixed time lag T=1, then the difference x(t,-1) =x(t)-x(t-1) has a stationary 1-point density but is generally not ergodic

(all higher order martingale densities are generally nonstationary). This is the generalization of so-called 'white noise' required for integration I(d).

The simplest example of a martingale where I(1) is impossible is defined by

$$dx = b(t)dB(t) \qquad (9)$$

because the increments are nonstationary unless b(t)=constant. But this increment nonstationarity can be easily eliminated, yielding an I(1) process, by discovering the time transformation that reduces (9) to the Wiener process [10,25]. The required transformation can easily be constructed once b(t) is known, and b(t) could be discovered *if* one could measure the time dependence of the process variance

$$\langle x^2(t) \rangle = \int_0^t b^2(s)ds. \qquad (10)$$

As an example, $b(t)=t^{2H-1}$ yields $\langle x^2(t) \rangle = t^{2H}\langle x^2(1) \rangle$. It's explained in [9] that level variances generally cannot be measured even if ensemble averages can be constructed from a single historic time series like an FX series.

Here's another example of a process where I(d) is impossible, but a different kind of transformation (using Ito's lemma) 'flattens' the nonstationarity of the increments. Consider the simplest variable diffusion process, the drift-free lognormal process

$$dp = pdB \qquad (11)$$

where B is the Wiener process. The first differences are martingale increments

$$p(t+T)-p(t)+\int_{t}^{t+T} p(s)dB(s) \qquad (12)$$

and are nonstationary,

$$\langle(p(t+T)-p(t))^2\rangle = \int_{t}^{t+T}\langle p^2(s)\rangle ds \qquad (13)$$

because $\langle p^2(t)\rangle = Ce^t$. No order of differencing can eliminate this nonstationarity for any time lag T>0. If we should instead take log increments $x(t,T)=\ln(p(t+T)/p(t))$, differences of log returns levels $x(t)=\ln p(t)$, then we would obtain the Wiener process with constant (removable) drift, so that with T=1 we could then construct a stationary 1-point density of increments via transformations. It would appear that I(1) is saved by coordinate transformations, indeed in this case it is saved, but our point is that this particular example is a pure mathematical accident: nonstationary increments generally cannot be eliminated by a coordinate transformation. Unlike the lognormal process, an arbitrary martingale is topologically inequivalent to the Wiener process [10]. However, for any stationary increment martingale I(1) is trivially possible with T=constant. *Stationary increment martingales define the class of pure noise processes that are I(1).*

Can we produce a nontrivial example of a stationary increment martingale? For any martingale, the mean square fluctuation is given by [10]

$$\langle x^2(t,T)\rangle = \langle x^2(t+T)\rangle - \langle x^2(t)\rangle \qquad (14)$$

and vanishes for a levels variance linear in t. If the diffusion coefficient is time translationally invariant, D(x,t)=D(x), then the process variance in linear in t. Also, if the diffusion coefficient scales with a Hurst exponent H=1/2, $D(x,t) = D(|x|/|t|^{1/2}, 0)$, then again the process variance is linear in time t. If we use $f_1(x,t) = p_2(x,t|0,0)$ in the increment density

$$f(z,t,t+T) = \int dx\, p_2(x+z,t+T|x,t) f_1(x,t) \qquad (15)$$

then we cannot prove that f is t-independent. Because of the argument x+z in the transition density under the integral sign, we cannot prove that we have stationary increments unless we would have spatial translational invariance, but that would restrict us to the Wiener process and financial returns are not a Wiener process (they also are not time translationally invariant, nor can they be shown to be scale invariant). So we cannot yet rigorously exhibit a nontrivial example of an I(1) process. The full condition of stationarity of increments is also impossible to verify or satisfy empirically. That is, 'true' I(1) is an unrealistic condition, empirically seen. Note that even if we can't prove stationary increments rigorously for the two separate cases of time-translational invariance and scaling of D(x,t), we can still use (15) to prove that the mean square fluctuation is t-independent. This may be as close as we can get, in practice, to increment stationarity.

Empirically seen, establishing stationarity of increments from a time series would generally be impossible because we usually cannot hope to discover the 1-point density of increments due to sparseness of the data. We therefore introduce the definition of 'weak stationarity of increments', requiring only that the mean square fluctuation is t-independent, $\langle x^2(t,T) \rangle = \langle x^2(0,T) \rangle$.

In empirical analyses or simulations in marcoeconomics where a series 'appears' visually to be I(1), we insist on asking: *what was measured, and how, to show or even indicate that the increments should be stationary?* Where cointegration has been used in macroeconomic analysis [26], we ask pointedly what was the empirical evidence that was produced by the practitioners to show that the random variables x(t) representing measurable macroeconomic variables are even approximately I(1)? According to Johansen [5], the evidence is often provided only by the visual inspection of a time series (see the graphs in [27], e.g.). Given the standard statistical tests including the search for a unit root, we suspect that the answer ranges from completely inadequate to none. *The unit root test is an insufficient test for a martingale, and even then does not test for increment stationarity*. For a martingale, it's both necessary and sufficient to test for a time lag T showing evidence of vanishing increment autocorrelations (fig. 1 in [9]). The equivalence of vanishing increment autocorrelations to a martingale condition has been explained elsewhere [9,10]. We've explained in [3,9] how to construct ensemble averages from a single, historic time series.

## 5.2 Nonstationary Increment Martingales and Finance Markets

Integration I(d) is in principle impossible, whether for T=1 period or for any value of T, if martingale noise increments are nonstationary, as indeed they are in intraday finance data. We detrended and analyzed [3] a 6 year Euro/Dollar Olsen & Assoc. returns series taken at T=10 min. intervals to eliminate increment autocorrelations (fig. 1). We thereby established that detrended FX returns $x(t)=\ln(p(t)/p_c)$ describe approximately a martingale for T≥10 min. of trading. The nonstationarity of intraday first differences for

T≥10min. is shown as fig. 1 in [3]. The scatter that figure is due to inadequate intraday statistics caused by the ensemble average required to handle nonstationary increments correctly. In a 6 year time series, there are only 1500 points for each intraday time t from which to calculate ensemble averages [3,9].

With nonstationary increments, the increment $x(t,-T) = x(t) - x(t-T)$ depends irreducibly on the starting time t and no amount of higher order differencing can eliminate this nonstationarity. That is, integration I(d) is impossible for intraday FX data, differencing cannot lead to stationarity for financial time series. This, and not the restriction to T=1, is the fundamental reason that 'long time equilibrium relations' due to ergodicity have not been observed via cointegration of financial variables like an FX rate and the relative prices levels in the two corresponding countries [2]. A visual inspection of the second fig. 1 in [3] indicates that the increments 'look' approximately stationary on the time scale of a day, but we have already explained above why no ergodicity follows from increment stationarity. Fig. 2 of [3] shows the region for which we were able to fit the data via a scaling model

$$D(x,t) = |t|^{2H-1}(1 + |x|/|t|^H)/H \qquad (16)$$

with H≈.35, and this model explicitly has nonstationary increments even in the weak sense if $H \neq 1/2$.

One may ask if tick data [28,29] yield stationary increments. A moment's reflection shows that the transformation from tick time to real time is not deterministic (given a tick, we cannot state with certainty when the next tick occurs)), so we cannot transform systematically from a real time analysis to a tick analysis. If the ticks are recorded in real time then a local time transformation from ticks to real time would be numerically possible. As a related fact, Gallucio et al [30] performed a local time transformation to obtain H≈1/2 numerically using a much shorter FX time series studied in real time, but there is no formula for transforming the entire series globally.

## 6. Conclusions

For martingale processes integration I(d) is restricted to stationary increment martingales. I(1) then covers all possibilities, I(d) for d≥2 is already included in I(1). There is no ergodicity nor is there a spectral decomposition for a stationary increment series x(t,-1). Most martingales, like detrended finance market processes for time lags T≥10 min. in intraday trading, have nonstationary increments. In general, a transformation to eliminate that nonstationarity cannot be constructed. These facts should change both econometrics and macroeconomic expectations dramatically. In particular, it would seem the height of folly to base policy decisions (monetary, regulatory, and taxation) on stationary predictions or expectations derived from regression analysis. In particular, the reason that equilibrum relationships cannot be seen empirically between FX rates and relative price levels is that there is no basis for long time averages to converge to ensemble averages, there is no empirically sound basis for assuming statistical equilibrium in markets.

## Acknowledgement

JMC is grateful to Barkley Rosser and Duncan Foley for helpful suggestions via email, to Giulio Bottazzi for hospitality and a lively discussion of i.i.d. noise and rational expectations in Pisa, to Steve Keen for ref. [29] and informative discussions of the money supply via email, and to Peter Reinhard Hansen for very helpful criticism. I'm also especially grateful to Søren Johansen for extremely helpful criticism, along with a cointegration tutorial that corrected technical mistakes in our description of cointegration and i.i.d. noise.